\documentclass[letterpaper, 10 pt, conference]{ieeeconf}  

\IEEEoverridecommandlockouts                              
\usepackage{cite}
\usepackage{amsmath}

\usepackage{comment}
\usepackage{graphicx}

\usepackage[utf8]{inputenc}
\usepackage[T1]{fontenc}
\usepackage{siunitx}
\usepackage{color}
\definecolor{martinsyellow}{RGB}{200,200,0}

\title{\LARGE \bf
Predicting EEG Responses to Attended Speech via Deep Neural Networks for Speech
}

 \author{Emina Alickovic, Tobias Dorszewski, Thomas U. Christiansen, Kasper Eskelund, \\ Leonardo Gizzi,  Martin A. Skoglund, Dorothea Wendt
\thanks{E.A., T.D.,  M.A.S. and D.W. are with Eriksholm Research Centre, Snekkersten, Denmark (e-mail: \{eali, todk, mnsk, dowe\}@eriksholm.com). E. A. and M.A.S. are also with the  Department of Electrical Engineering, Linköping University, Linkoping, Sweden (e-mail: \{firstname.lastname\}@liu.se). T.D. and L.G. are with the Institute for Modelling and Simulation of Biomechanical Systems, University of Stuttgart, Germany (e-mail: \{firstname.lastname\}@imsb.uni-stuttgart.de). D.W. is also with the Department of Health Technology, Technical University of Denmark, Lyngby, Denmark. K.E. and T.U.C. are with Oticon A/S, Smørum, Denmark (e-mail: \{thch, ksee\}@demant.com).}
\thanks{The data analysis in this paper is based on a Master's Thesis work \cite{tobias_thesis}.}
\thanks{This work has been submitted to the IEEE for possible publication. Copyright may be transferred without notice, after which this version may no longer be accessible.}
}
\begin{document}

\maketitle
\thispagestyle{empty}
\pagestyle{empty}

\begin{abstract}

Attending  to the speech stream of interest in multi-talker environments can be a challenging task, particularly for listeners with hearing impairment. Research suggests that neural responses assessed with electroencephalography (EEG) are modulated by listener’s auditory attention, revealing selective neural tracking (NT) of the attended speech. 
NT methods 
mostly rely on hand-engineered acoustic and linguistic speech features 
to predict the neural response. 
Only recently, 
 deep neural network (DNN) models without specific linguistic 
 information have been used to extract speech features for NT, demonstrating that speech features in hierarchical DNN layers  can predict neural responses throughout the auditory pathway. 
In this study, we go one step further to investigate the suitability of similar DNN models for speech to predict neural responses to competing speech observed in EEG. 
We recorded EEG data using a 64-channel acquisition system from 17 listeners with normal hearing 
instructed to attend to one of two competing talkers. 
Our data revealed that EEG responses are significantly better predicted by  DNN-extracted speech features than by hand-engineered acoustic features. Furthermore, analysis of hierarchical DNN layers showed that early layers yielded the highest predictions. Moreover, we found a significant increase in auditory attention classification accuracies with the use of  DNN-extracted speech features over the use of hand-engineered acoustic features.  
These findings open a new avenue for development of new NT measures to 
evaluate and further advance hearing technology.


\end{abstract}

\section{INTRODUCTION}

The challenge of attending the target speech signal while ignoring other sounds is an extensively-studied problem known as the “cocktail party” problem \cite{cherry1953some}. To date, research on "cocktail party" environments has progressed significantly to a point where it is now possible to decode (i.e., classify) the attended speech  by quantifying “neural tracking" (NT) of speech \cite{alickovic2019tutorial, geirnaert2021electroencephalography}, i.e.,  by comparing the neural activity of the listener, recorded with electroencephalogram (EEG), to the activity of multiple candidate speech sources in a listening environment \cite{o2015attentional,alickovic2020neural}. NT methods have allowed to assess the real benefits of the hearing aids  \cite{alickovic2020neural, alickovic2021effects}.

NT methods  involve encoding of the speech by estimating the temporal response function (TRF) that linearly maps time-lagged speech signals to EEG. 
NT methods proposed in the literature mostly rely on hand-engineered acoustic and linguistic speech feature that include for example  speech envelope, spectrogram,  pitch, phonetic and lexical features \cite{biesmans2016auditory, di2015low, broderick2018electrophysiological, gillis2021neural, patel2022interaction}. 
Despite the successful usage of acoustic-linguistic features in NT, it remains unclear to what extent language models \cite{baevski2020wav2vec, hsu2021hubert} may be used to extract relevant features. 

Modern artificial intelligence models using deep neural networks (DNN) revolutionized the field of speech representation showing that DNN models, without specific language knowledge, can derive speech features that correlates well with neural responses  recorded with functional magnetic resonance imaging (fMRI), magnetoencephalography (MEG) and electrocorticography (ECoG) \cite{goldstein2022shared, caucheteux2022brains, li2022dissecting}. Furthermore, recent work has suggested that DNN models  can successfully predict speech comprehension from the neural responses recorded with  fMRI \cite{caucheteux2022deep}. 
Lastly, Li et. al. \cite{li2022dissecting} demonstrated that speech features in hierarchical DNN layers can better predict neural responses than hand-engineered  acoustic-phonetic features, as observed in high signal-to-noise (SNR) ECoG signals, throughout the auditory pathway \cite{li2022dissecting}.  However, whether DNN-derived speech features can better predict noninvasive, low SNR EEG responses and, classify attention in  multi-talker listening environments remains unknown. 

To address this issue, we compare a wide variety of hand-engineered acoustic and DNN-derived speech features in light of  human neural responses to competing speech. Specifically, we analyze the EEG responses of 17 healthy  younger adults. During 45 min-long sessions the listeners were instructed to attend to one of two competing talkers. We trained a variety of NT models and compare their ability to linearly map speech onto the EEG recordings. Furthermore, we investigate the hierarchy of layers in the  DNN models.
Finally, we trained a variety of NT models and
 compare their ability classify the attended speech. 
\vspace{-2pt}
\section{METHODS} 
The experimental protocol was reviewed and approved by the ethics committee for the capital region of Denmark (journal number H-21065001). The study was conducted according to the Declaration of Helsinki, and all the participants gave a written 
consent prior to the experiment.

\subsection{Study Design}
\subsubsection{Participants} Participants comprised 17 younger adults (9 males, mean age 29.0, SD 6.4). All participants were native Danish speakers, had normal or corrected-to-normal vision, had no history of neurological disorders, dyslexia, or diabetes mellitus, and had clinically normal hearing.

\subsubsection{Stimuli and Recording} 
During the experimental session presented in this study, participants were asked to attend to one of two different audiobooks narrated in Danish. 
The stimuli comprised 33 $\sim$\SI{1}{\minute}-long segments from audiobook recordings of \textit{Himalaya i sigte} (a story of traveling in the Himalayas) read by a female talker and \textit{Simon} (a biography on Simon Spies) read by a male talker, and sampled at \SI{44.1}{\kilo\hertz}.
Prolonged silent periods in the speech stimuli longer than \SI{500}{\milli \second} were shortened to \SI{500}{\milli \second}. Stimuli were routed through a sound card (RME Hammerfall DSP multiface II, Audio AG,  Germany) and were played via loudspeakers (Genelec 8040A; Genelec Oy, Finland) at an average intensity of 70 dB SPL each positioned $\pm 30^{\circ}$ to the left or right of the center. EEG data were acquired at a sampling rate of \SI{8192}{\hertz} with a BioSemi ActiveTwo 64-channel EEG recording system in 10-20 layout. 

\subsubsection{Experimental Session Design} A total of 33 trials were  conducted, with the first trial used for training and 32 trials used for analysis. Each trial consisted of \SI{5}{\second} of silence, \SI{1}{\minute} of speech mixture and two 2-choice questions on attended story to keep the participants alert. 
The 32 trials were divided into 8 blocks of 4 randomized consecutive trials, with 2 blocks for each of “male right”, “male left”, “female right” and “female left”. Before each block, a visual cue on the screen and \SI{5}{second} of the to-be-attended speech  were presented indicating talker (male or female) and the side (left or right) to be attended. 

\vspace{-4pt}

\subsection{Neural Data Analysis}
\subsubsection{EEG Preprocessing} The EEG data were re-referenced to the average of the mastoid electrodes, band-pass filtered between 0.1 and \SI{10}{\hertz} and re-sampled to \SI{100}{\hertz}. Subsequently, signal components of non-neural origin were removed using a procedure based on independent component analysis \cite{bell1995information}. Next, the data were  filtered between 1 and \SI{10}{\hertz} and normalized to zero mean and unit variance. In a final step,  data were segmented into trials of \SI{59}{\second} duration from 0 to \SI{59}{\second} relative to the onset of the speech.

\subsubsection{Neural Tracking of Speech}
\paragraph{Quantifying Brain Prediction Scores}
The TRF framework allows to study how the brain processes competing speech. It includes two stages: a training stage to derive TRFs for the each talker and a testing stage to quantify  how well EEG responses can be predicted. In the training stage,  time-lagged speech features of each talker $S$ are linearly mapped to the EEG response(s) $R$ of the listener 
based on TRF $W$ derived  via regularized linear regression (rLR) with a parameter $\lambda$ to control for overfitting \cite{alickovic2019tutorial}: 
\begin{equation}\label{linear with delay}
    W = \left[\mathcal{H}(S)^T\mathcal{H}(S)+\lambda I\right]^{-1}\mathcal{H}(S)R \
\end{equation}
where $\mathcal{H}(*)$ is a Hankel matrix (see \cite{alickovic2019tutorial} for more details). In the testing stage, EEG responses are predicted as: 
\begin{equation}
    \widehat{R} = \mathcal{H}(S) W 
\end{equation}
 The quality of the prediction is  quantified in terms of a brain prediction score (BPS) measuring the correlation (a Pearson's $r$) between the true  and reconstructed  EEG responses.
 
\paragraph{Classifying Auditory Attention} The NT framework allows to classify auditory attention (i.e., identify the attended speech) in multi-talker environments. 
In order to classify which of the streams a listener attended to, two TRF models $W_{att}$ and $W_{ign}$ from \eqref{linear with delay} are assembled to 
become two competing prediction models for every single EEG channel. Next, the EEG signals ($\widehat{R}_{att}$ and $\widehat{R}_{ign}$) are independently predicted from the attended ($S_{att}$) and the ignored ($S_{ign}$) speech signals. To estimate which of the predicted EEG signals (${{\widehat{R}}_{att}}$  versus ${{\widehat{R}}_{ign}}$ ) are most likely representing the attended speech, we compute channel-by-channel brain prediction scores $BPS_{att}$ and  $BPS_{ign}$. Finally, we compare  $BPS_{att}$ and  $BPS_{ign}$ values averaged across all EEG channels and the signal with the highest BPS is classified as the attended speech.

\subsection{Speech Feature Extraction}

\subsubsection{Hand-Engineered Speech Features}
Hand-engineered acoustic features considered for this study included \textit{speech envelope} (the root-mean-square of \SI{10}{\milli\second} windows and scaled by raising the value to the power of a compression parameter of 0.3 or 1 indicating no compression), the \textit{envelope derivative}, 
\textit{spectrogram} (100 linearly spaced components between \SI{0}{\hertz} and \SI{8}{\kilo\hertz}  computed with a short-time Fourier transform at \SI{100}{\hertz}, and scaled by a compression parameter of 0.3), Mel Frequency Cepstral Coefficients (\textit{MFCC}) (13 components representing different frequencies between \SI{20}{\hertz} and \SI{8}{\kilo\hertz}, as well as their derivatives and the second derivative for a total of 39 features) and \textit{pitch}
(absolute and relative pitch, and pitch change computed as in \cite{li2022dissecting}). The MFCC feature set was also used as an initialization of the labels for DNN training. Similar to \cite{li2022dissecting}, we included a baseline model comprising full acoustic features. 

\subsubsection{DNN-extracted Speech Features}
We employ the Hidden unit BERT (HuBERT) DNN model - a transformer-based self-supervised model for speech feature learning \cite{hsu2021hubert}. A major component of HuBERT model training is applying the predictive loss over the masked portions of speech driving the model to learn a fused acoustic and language feature set over the speech input. 
Using  ECoG recordings, it has recently been shown that the HuBERT DNN model yielded the best BPS among the benchmarks DNN models \cite{li2022dissecting}. 

Our goal is to extract relevant speech features in different DNN layers in order to use them  as inputs to the NT models. 
With the speech material presented in our study being in Danish, different methods are employed to obtain a Danish version of HuBERT DNN model. For a complete description please refer to \cite{tobias_thesis}.

Since both the audio used in the experiments and the LibriSpeech corpus  \cite{panayotov2015librispeech} of the English HuBERT model stem from audiobooks, we collected a similar Danish speech corpus for training purposes. Only continuous, clearly spoken, Danish speech without background noise was included. The main source was speech materials used in previous studies conducted at Eriksholm Research Centre. Additionally, the publicly available Danish audiobooks found on LibriVox \cite{kearns2014librivox} were included. For all speech files, the starts of the files with an introduction (e.g.,  the name of the reader or the book title) were removed. Mono channel audio signal was created by averaging the stereo channels, resampled to \SI{16}{\kilo\hertz} and divided into equally long segments. To avoid a strong effect of individual speakers, the amount of speech data was limited to 300 files per speaker. In total 3900 such audio files ($>$ \SI{65}{\hour})  were prepared and divided into training and validation sets in a common $80/20$ split.

An important component of the HuBERT DNN models are the artificially created labels with which the DNN is trained. Based on MFCCs, all speech segments were clustered with k-means. These clusters were used as labels for the DNN training. We considered three major HuBERT DNN models, see Table \ref{tab:DNNs}. First, the original 'base' HuBERT DNN model, referred to as the 'English DNN' in this study, was trained on \SI{960}{\hour} of continuous English speech from the LibriSpeech corpus \cite{hsu2021hubert}. 
Second, the Danish HuBERT DNN model, here referred to as the 'Danish DNN', used  the \SI{\sim 65}{\hour} data set of Danish speech for training with randomly initialized weights which mimics the original training of the  base English DNN \cite{hsu2021hubert}.
Third, a self-supervised fine-tuning (SSFT) was added as an additional training option. The  English DNN is used to initialize the network weights. Here, only the latest layers of the pre-trained DNN are replaced with new layers to allow prediction of different clusters. The training objective during SSFT is the HuBERT objective of classifying segments of the unseen speech data. We refer to this model as SSFT A1.




\begin{table}
\caption{Overview of the trained HuBERT DNN models. L9 denotes layer 9 in DNN HuBERT model. }
\begin{center}
\resizebox{\columnwidth}{!}{\begin{tabular}{ l | | l | l | l}
DNN Name & Data set & Weight source & Labels  \\
\hline
\hline
English  & 960 h English & Random & 100, from MFCC \\
\hline
Danish & 65 h Danish & Random & 100, from MFCC \\
\hline
SSFT A1 & 65 h Danish & English  & 250, from L9 of Danish \\
\hline
\end{tabular}}
\end{center}
\label{tab:DNNs}
\vspace{-.5cm}
\end{table}

\section{RESULTS \& DISCUSSION}
\subsection{Predicting EEG from DNN-Extracted Speech Features}
We first test whether DNN-extracted speech features linearly predict  EEG responses. To this aim, we fit a rLR to predict the EEG activity elicited by the attended speech from the HuBERT model input with the same speech. We then compute a BPS, i.e. the correlation between the true EEG responses and the EEG responses predicted from the rLR. 
On average across EEG channels, the BPS for an English DNN HuBERT model  are significantly distributed above zero with the mean BPS of 0.051 at layer 1 (L1), 0.054 at layer 5 (L5) and 0.05 at layer 12 (L12). Using speech features from the Danish DNN HuBERT model, a lower BPS of 0.035 at L1 to 0.015 at L12 were achieved. 

\begin{figure}[h]
    \centering
    \includegraphics[width=\columnwidth]{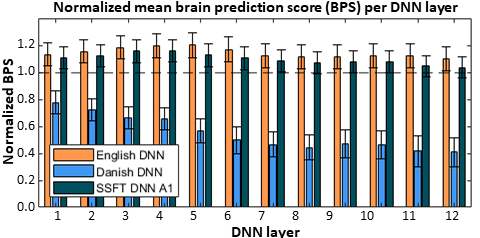}
    \caption{The normalized mean brain prediction scores (BPS) for predicting the EEG activity elicited by the attended speech from the HuBERT model input with the same speech. 
    A BPS higher than 1 signifies a higher quality of prediction from DNN-extracted speech features than from the hand-engineered acoustic features. 
    }
    \label{fig:dnn_eeg_bps}
\end{figure}

\begin{figure}[h]
    \centering
    \includegraphics[width=\columnwidth]{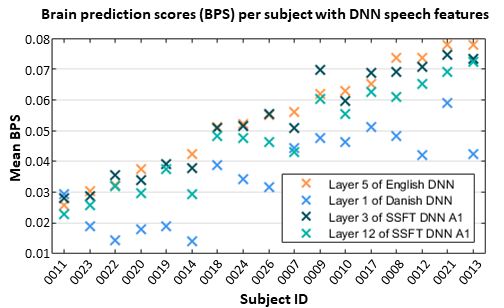}
    \caption{ The mean BPS for each subject predicting the EEG responses from the HuBERT DNN models averaged across all trials and channels. 
    }
    \label{fig: dnn_eeg subjects}
\end{figure}

Second, we evaluate whether DNN-extracted speech features can better predict EEG responses than hand-engineered acoustic features (see Fig. \ref{fig:dnn_eeg_bps}-\ref{fig: dnn_eeg subjects}). Similar to the speech from HuBERT model, we first fit a rLR to predict the EEG activity from the acoustic features. On average, a BPS of 0.039,  0.042, 0.029, 0.037, 0.032 and 0.038  for acoustic envelope, all envelope features, pitch, MFCC, spectrogram, and all features, respectively, were observed. Next, we compute a normalized BPS, i.e., the squared BPS from the DNN HuBERT model divided by the squared BPS with all envelope features as input to the rLR.  Lastly, we observe that both the English DNN and the SSFT DNN A1 HuBERT models provide normalized BPS scores that are significantly higher than 1 at their best layer (p $<$ 0.05, 2-sided t-tests with BPS for 64 EEG channels). Our results are consistent with previous findings suggesting that DNN-derived speech features correlates well with the neural activity \cite{goldstein2022shared, caucheteux2022brains, li2022dissecting}. 

\vspace{-.2cm}

Third, analysis of hierarchical DNN layers shows that early layers (layer group 1: L1-L5) yielded higher BPS than later layers (layer group 2: L6-L12; $p<0.05$ for one-way ANOVA factor 'layer group'). 
This is in line with recent studies  providing evidence for the  speech processing hierarchy within auditory cortex  \cite{kell2018task, millet2021inductive}. 

\vspace{-4pt}
\vspace{-.4cm}

\subsection{Attention Classification with DNN-extracted Speech }
\begin{figure}[h]
    \centering
    \includegraphics[width=\columnwidth]{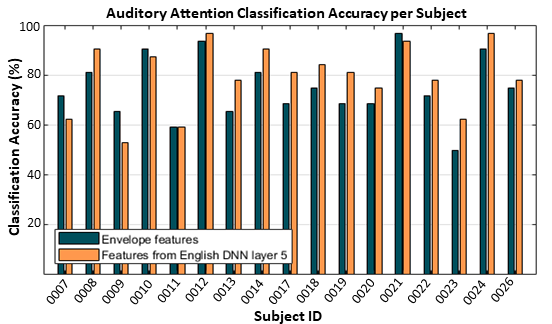}
    \caption{Auditory attention classification results per subject 
    shown for the predictions based on the best acoustic feature set and the best DNN feature set (features from the layer 5 of an DNN trained only on English audio).}
    \label{fig: attention-decoding}
\end{figure}

Classifying auditory attention is notoriously challenging \cite{geirnaert2021electroencephalography}. This issue poses strong
limitations on the future application of NT methods to hearing devices.  While hand-engineered acoustic features can be used in NT methods to decode attention, we show that DNN-extracted speech features yield results that are consistently higher than those described in neuroscientific literature (see Fig. \ref{fig: attention-decoding}). We find a significant increase in classification accuracy with the use of  DNN-extracted  features over the use of acoustic features. 

For the best acoustic feature set (envelope features), a mean attention (attended vs. ignored speech) classification  accuracy of 75\% (SD 43\%) was achieved. With the best DNN feature set (from the L5 of the English DNN), the attention classification accuracy improved to 79\% (SD 40\%), yielding  statistically significant differences (p = 0.0319, 2-sided t-test with 17 subjects). 

\vspace{-4pt}

\vspace{-4pt}







\section{CONCLUSIONS}
We propose a new framework to predict EEG responses to attended speech. 
Overall, the present study suggests  DNN models for speech can retrieve information that correlate to speech processing hierarchy. Interestingly, our analyses highlights that EEG responses are significantly better predicted by DNN-extracted speech features than by hand-engineered acoustic features. Furthermore, analysis of hierarchical DNN layers shows that early layers yield the highest predictions. Finally, we find a significant increase in auditory attention classification accuracy with the use of DNN-extracted speech features over the use of hand-engineered acoustic features. 
In sum, 
NT methods could used to evaluate and further advance hearing technology and we propose a new approach to increase EEG prediction and attention classification accuracy.

\addtolength{\textheight}{-12cm}   





\bibliography{arXiv.bib}
\bibliographystyle{IEEEtran}

\end{document}